\documentclass[12pt]{article}
\usepackage{epsf}
\title{ {\bf
Lepton flavor violating $Z\rightarrow l_1^+ l_2^-$ decay in the general two Higgs 
Doublet model}}
\author{\vspace{1cm}\\
        {\bf E. O. Iltan}
        \thanks{E-mail address:
        eiltan@heraklit.physics.metu.edu.tr}\,\,\, and 
   {\bf I. Turan}
        \thanks{E-mail address:
        ituran@.metu.edu.tr}
 \\
        Physics Department, Middle East Technical University \\
        Ankara, Turkey\\}

\date{}

\begin{document}
\setlength{\baselineskip}{24pt}
\maketitle
\setlength{\baselineskip}{7mm}
\begin{abstract}
We calculate lepton flavor violating $Z\rightarrow l^+ l^-$ decay in the
framework of the general two Higgs Doublet model. In our calculations we
used the constraints for the Yukawa couplings $\bar{\xi}^{D}_{N,\tau e}$ 
and $\bar{\xi}^{D}_{N,\tau\mu}$ coming from the experimental result of muon 
electric dipole moment and upper limit of the  $BR(\mu\rightarrow e\gamma)$. 
We observe that it is possible to reach the present experimental upper
limits for the branching ratios of such Z decays in the model III. 
\end{abstract} 
\thispagestyle{empty}
\newpage
\setcounter{page}{1}
\section{Introduction}
Lepton Flavor Violating (LFV) interactions reached great interest 
since the related experimental measurements are improved at present. Among
them, the lepton flavor changes in Z decays, such as 
$Z\rightarrow e \mu$, $Z\rightarrow e \tau$ and $Z\rightarrow \mu \tau$ are 
important for the search of neutrinos, their mixing and possible masses, and 
the physics beyond the Standard model (SM). With the Giga-Z option of the
Tesla project, the production of Z bosons at resonance is expected to
increase \cite{Hawkings} and this forces to study on such Z decays more
precisely. 

Since the lepton flavor is conserved in the SM, one needs an extended theory
to describe the lepton flavor violating Z decays. One of the possibility is
the extension of the SM, so called $\nu$SM, by taking neutrinos massive and 
permitting the lepton mixing mechanism \cite{Pontecorvo}. In this case the 
lepton sector is analogous to the quark sector. Considering the branching 
ratio ($BR$) 
\begin{eqnarray}
BR(Z\rightarrow l_1^{\pm} l_2^{\pm})=\frac{\Gamma (Z\rightarrow \bar{l}_1 l_2+
\bar{l}_2 l_1)}{\Gamma_Z} \, , \nonumber 
\end{eqnarray}
the first predictions for such Z decays are given in \cite{Riemann,Ganapathi}. 
The best experimental limits obtained at LEP 1 \cite{PartData} are 
\begin{eqnarray}
BR(Z\rightarrow e^{\pm} \mu^{\pm}) &<& 1.7\times 10^{-6} \,\,\, \cite{Opal} 
\nonumber \, , \\
BR(Z\rightarrow e^{\pm} \tau^{\pm}) &<& 9.8\times 10^{-6}\,\,\,
\cite{Opal,L3} \nonumber \, , \\ 
BR(Z\rightarrow \mu^{\pm} \tau^{\pm}) &<& 1.2\times 10^{-5} \,\,\,
\cite{Opal,Delphi} 
\label{Expr1}
\end{eqnarray}
and with the improved sensitivities at Giga-Z \cite{Wilson} these numbers 
could be pulled down to 
\begin{eqnarray}
BR(Z\rightarrow e^{\pm} \mu^{\pm}) &<& 2\times 10^{-9}  \nonumber \, , \\
BR(Z\rightarrow e^{\pm} \tau^{\pm}) &<& f\times 6.5\times 10^{-8}
\nonumber \, , \\ 
BR(Z\rightarrow \mu^{\pm} \tau^{\pm}) &<& f\times 2.2\times 10^{-8}  
\label{Expr2}
\end{eqnarray}
with $f=0.2-1.0$.
In the framework of $\nu$SM with light neutrinos the theoretical prediction 
is extremely small and far from these limits \cite{Riemann,Illana}    
\begin{eqnarray}
BR(Z\rightarrow e^{\pm} \mu^{\pm})\sim BR(Z\rightarrow e^{\pm} \tau^{\pm}) 
&\sim& 10^{-54}  \nonumber \, , \\
BR(Z\rightarrow \mu^{\pm} \tau^{\pm}) &<& 4\times 10^{-60}  
\label{Theo1}
\end{eqnarray}
Another possiblity to increase the corresponding $BR$ is the extension 
of the $\nu$SM with one heavy ordinary Dirac neutrino \cite{Illana}.
Heavy neutrinos are expected in string-inspired models \cite{Witten} and 
some GUTs \cite{Langacker}. In the $\nu$SM with one heavy neutrino, it is 
necessary to include a heavy charged lepton in the theory. In the famework 
of this model, its is possible to observe their effects from Z decays if the 
neutrinos with a mass of several hundered GeV exist. Further scenario to 
increase the $BR$ is the $\nu$SM extended with two heavy right-handed singlet 
Majorana neutrinos \cite{Illana}. In this case, it is possible to reach the 
experimental upper limits in the large neutrino mass region. Lepton flavor 
Violating Z decays are also studied in the framework of the Zee model 
\cite{Ghosal}. According to this work, among all three lepton flavor violating 
decay modes of Z, only $Z\rightarrow e\tau$ decay has the largest 
contribution which is less than the present limits. Other two decays are small 
to be observed in the next linear colliders. 

In our work, we study $Z\rightarrow e^{\pm} \mu^{\pm}$, $Z\rightarrow 
e^{\pm} \tau^{\pm}$ and $Z\rightarrow \mu^{\pm} \tau^{\pm}$ decays in the 
model III version of 2HDM, which is the minimal extension of the SM. Since 
there is no CKM type matrix and therefore no charged FC interaction in the 
leptonic sector according to our assumption, the source of lepton flavor 
violating  Z decays are the neutral Higgs bosons $h^0$ and $A^0$ with the 
Yukawa couplings which allow tree level flavor changing neutral currents 
(FCNC). The choice of complex Yukawa couplings brings the possibility of 
non-zero electric dipole moments (EDM) of leptons and this ensures to 
restrict the Yukawa couplings using the present experimental limits. Further 
the lepton flavor violating interaction $\mu\rightarrow e \gamma$ is possible 
in this model and can be used to predict the constraint for the Yukawa 
couplings (see \cite{ErLFV}). Calculations are done in one loop level and 
it is shown that the experimental upper limits for Z decays underconsideration 
can be reached by playing with the free parameters of the model III respecting 
the above restrictions.  

The paper is organized as follows:
In Section 2, we present the explicit expressions for the Branching ratios
of $Z\rightarrow e^- \mu^+$, $Z\rightarrow e^- \tau^+$ and 
$Z\rightarrow \mu^- \tau^+$ in the framework of the model III. Section 3 is 
devoted to discussion and our conclusions.
\section{$Z\rightarrow l_1^- l_2^+$ decay in the general two Higgs Doublet 
model.} 
In the SM and the model I and II version of 2HDM the FCNC at tree level is 
forbidden. However model III version of 2HDM permits  FCNC interactions at 
tree level and makes flavor violating interactions possible. With the choice 
of complex Yukawa couplings, the CP violation can also exist and this leads 
to appearence of non-zero EDM of fermions. The Yukawa interaction for the 
leptonic sector in the model III is
\begin{eqnarray}
{\cal{L}}_{Y}=
\eta^{D}_{ij} \bar{l}_{i L} \phi_{1} E_{j R}+
\xi^{D}_{ij} \bar{l}_{i L} \phi_{2} E_{j R} + h.c. \,\,\, ,
\label{lagrangian}
\end{eqnarray}
where $i,j$ are family indices of leptons, $L$ and $R$ denote chiral 
projections $L(R)=1/2(1\mp \gamma_5)$, $\phi_{i}$ for $i=1,2$, are the 
two scalar doublets, $l_{i L}$ and $E_{j R}$ are lepton doublets and
singlets respectively. Here $\phi_{1}$ and $\phi_{2}$ are chosen as
\begin{eqnarray}
\phi_{1}=\frac{1}{\sqrt{2}}\left[\left(\begin{array}{c c} 
0\\v+H^{0}\end{array}\right)\; + \left(\begin{array}{c c} 
\sqrt{2} \chi^{+}\\ i \chi^{0}\end{array}\right) \right]\, ; 
\phi_{2}=\frac{1}{\sqrt{2}}\left(\begin{array}{c c} 
\sqrt{2} H^{+}\\ H_1+i H_2 \end{array}\right) \,\, ,
\label{choice}
\end{eqnarray}
and the vacuum expectation values are  
\begin{eqnarray}
<\phi_{1}>=\frac{1}{\sqrt{2}}\left(\begin{array}{c c} 
0\\v\end{array}\right) \,  \, ; 
<\phi_{2}>=0 \,\, .
\label{choice2}
\end{eqnarray}
This choice ensures that the SM particles are collected in the first doublet 
and the ones beyond in the second doublet. The part which produce FCNC at 
tree level is  
\begin{eqnarray}
{\cal{L}}_{Y,FC}=
\xi^{D}_{ij} \bar{l}_{i L} \phi_{2} E_{j R} + h.c. \,\, .
\label{lagrangianFC}
\end{eqnarray}
Here the Yukawa matrices $\xi^{D}_{ij}$ have in general complex entries. 
Note that in the following we replace $\xi^{D}$ with $\xi^{D}_{N}$ where 
"N" denotes the word "neutral" and define $\bar{\xi}^{D}_N$ which satisfies
the equation $\xi^{D}_N=\sqrt{\frac{4\,G_F}{\sqrt{2}}}\, \bar{\xi}^{D}_N$. 
In our analysis, we take $H_1$ and $H_{2}$ (see eq. (\ref{choice})) as the 
mass eigenstates $h^0$ and $A^0$ respectively since no mixing between
CP-even neutral Higgs bosons $h^0$ and the SM one, $H^0$, occurs at tree 
level. 

The general effective vertex for the interaction of on-shell Z-boson with a
fermionic current is given by
\begin{eqnarray}
\Gamma_{\mu}=\gamma_{\mu}(f_V-f_A\ \gamma_5)+ \frac{i}{m_W}\,(f_M+f_E\,
\gamma_5)\, \sigma_{\mu\,\nu}\, q^{\nu}
\label{vertex}
\end{eqnarray}
where $q$ is the momentum transfer, $q^2=(p-p')^2$, $f_V$ ($f_A$) is vector 
(axial-vector) coupling, $f_M$ ($f_E$) magnetic (electric) transitions of 
unlike fermions. Here $p$ ($-p^{\prime}$) is the four momentum vector of
lepton (anti-lepton). The necessary 1-loop diagrams due to neutral Higgs 
particles are given in Fig. \ref{fig1ver}.  

Taking into account all the masses of internal leptons and external
lepton (anti-lepton), the explicit expressions for $f_V$, $f_A$, $f_M$ and 
$f_E$ read as
\begin{eqnarray}
f_V&=& \frac{g}{64\,\pi^2\,cos\,\theta_W}
\int_0^1\, dx \,  \frac{1}{m^2_{l_2^+}-m^2_{l_1^-}}
\Bigg \{ c_V \, (m_{l_2^+}+m_{l_1^-})  
\nonumber \\ 
&\Bigg(&
(-m_i \, \eta^+_i + m_{l_1^-} (-1+x)\, \eta_i^V)\, ln \, \frac{L^{self}_
{1,\,h^0}}{\mu^2}+
(m_i \, \eta^+_i - m_{l_2^+} (-1+x)\, \eta_i^V)\, ln \, \frac{L^{self}_{2,\,
h^0}}{\mu^2}
\nonumber \\ &+&
(m_i \, \eta^+_i + m_{l_1^-} (-1+x)\, \eta_i^V)\, ln \, \frac{L^{self}_{1,\,
A^0}}{\mu^2} 
- (m_i \, \eta^+_i + m_{l_2^+} (-1+x) \,\eta_i^V)\, ln \, \frac{L^{self}_{2,\,
A^0}}{\mu^2} 
\Bigg)
\nonumber \\ &+&
c_A \, (m_{l_2^+}-m_{l_1^-}) \nonumber \\
&\Bigg ( & 
(-m_i \, \eta^-_i + m_{l_1^-} (-1+x)\, \eta_i^A)\, ln \, \frac{L^{self}_{1,\,
h^0}}{\mu^2} + 
(m_i \, \eta^-_i + m_{l_2^+} (-1+x)\, \eta_i^A)\, ln \, \frac{L^{self}_{2,\,
h^0}}{\mu^2}
\nonumber \\ &+&
(m_i \, \eta^-_i + m_{l_1^-} (-1+x)\, \eta_i^A)\, ln \, \frac{L^{self}_{1,\,
A^0}}{\mu^2} + 
(-m_i \, \eta^-_i + m_{l_2^+} (-1+x)\, \eta_i^A)\, ln \, \frac{L^{self}_{2,\,
A^0}}{\mu^2} 
\Bigg) \Bigg \}
\nonumber \\ &-&
\frac{g}{64\,\pi^2\,cos\,\theta_W} 
\int_0^1\,dx\, \int_0^{1-x} \, dy \, 
\Bigg \{
m_i^2 \,(c_A\, \eta_i^A-c_V\,\eta_i^V)\,(\frac{1}{L^{ver}_{A^0}}+
\frac{1}{L^{ver}_{h^0}}) 
\nonumber \\ &-&
(1-x-y)\,m_i\, 
\Bigg( 
c_A\,  (m_{l_2^+}-m_{l_1^-})\, \eta_i^- \,(\frac{1}{L^{ver}_{h^0}} 
- \frac{1}{L^{ver}_{A^0}})+ 
c_V\, (m_{l_2^+}+m_{l_1^-})\, \eta_i^+ \, (\frac{1}{L^{ver}_{h^0}} 
+ \frac{1}{L^{ver}_{A^0}}) \Bigg) 
\nonumber \\ &-&
(c_A\, \eta_i^A+c_V\,\eta_i^V)
\Bigg ( 
-2+(q^2\,x\,y+m_{l_1^-}\,m_{l_2^+}\, (-1+x+y)^2)\, (\frac{1}{L^{ver}_{h^0}} 
+ \frac{1}{L^{ver}_{A^0}})-ln\,\frac{L^{ver}_{h^0}}{\mu^2}\,
\frac{L^{ver}_{A^0}}{\mu^2}
\Bigg )
\nonumber \\ &-&
(m_{l_2^+}+m_{l_1^-})\, (1-x-y)\, 
\Bigg (
\frac{\eta_i^A\,(x\,m_{l_1^-} +y\,m_{l_2^+})+m_i\,\eta_i^-}
{2\,L^{ver}_{A^0\,h^0}}+\frac{\eta_i^A\,(x\,m_{l_1^-} +y\,m_{l_2^+})-
m_i\,\eta_i^-}{2\,L^{ver}_{h^0\,A^0}} \Bigg )
\nonumber \\ &+&
\frac{1}{2}\eta_i^A\, ln\,\frac{L^{ver}_{A^0\,h^0}}{\mu^2}\,
\frac{L^{ver}_{h^0\,A^0}}{\mu^2} 
\Bigg \}\,, \nonumber \\
f_A&=& \frac{-g}{64\,\pi^2\,cos\,\theta_W}
\int_0^1\, dx \,  \frac{1}{m^2_{l_2^+}-m^2_{l_1^-}}
\Bigg \{ c_V \, (m_{l_2^+}-m_{l_1^-})  
\nonumber \\ 
&\Bigg(&
(m_i \, \eta^-_i + m_{l_1^-} (-1+x)\, \eta_i^A)\, ln \, 
\frac{L^{self}_{1,\,A^0}}{\mu^2}
+
(-m_i \, \eta^-_i + m_{l_2^+} (-1+x)\, \eta_i^A)\, ln \, \frac{L^{self}_
{2,\,A^0}}{\mu^2}
\nonumber \\ &+&
(-m_i \, \eta^-_i + m_{l_1^-} (-1+x)\, \eta_i^A)\, ln \, \frac{L^{self}_{1,\,
h^0}}{\mu^2}+ (m_i \, \eta^-_i + m_{l_2^+} (-1+x)\, \eta_i^A)\, 
ln \, \frac{L^{self}_{2,\,h^0}}{\mu^2} \Bigg)
\nonumber \\ &+&
c_A \, (m_{l_2^+}+m_{l_1^-}) \nonumber \\
&\Bigg(& 
(m_i \, \eta^+_i + m_{l_1^-} (-1+x)\, \eta_i^V)\, ln \, \frac{L^{self}_{1,\,
A^0}}{\mu^2}- (m_i \, \eta^+_i + m_{l_2^+} (-1+x)\, \eta_i^V)\, 
ln \, \frac{L^{self}_{2,\,A^0}}{\mu^2}
\nonumber \\ &+&
(-m_i \, \eta^+_i + m_{l_1^-} (-1+x)\, \eta_i^V)\, ln \, \frac{L^{self}_{1,\,
h^0}}{\mu^2} + (m_i \, \eta^+_i - m_{l_2^+} (-1+x)\, \eta_i^V)\, 
\frac{ln \, L^{self}_{2,\,h^0}}{\mu^2}
\Bigg) \Bigg \}
\nonumber \\ &+&
\frac{g}{64\,\pi^2\,cos\,\theta_W} 
\int_0^1\,dx\, \int_0^{1-x} \, dy \, 
\Bigg \{
m_i^2 \,(c_V\, \eta_i^A-c_A\,\eta_i^V)\,(\frac{1}{L^{ver}_{A^0}}+
\frac{1}{L^{ver}_{h^0}}) 
\nonumber \\ &-&
m_i\, (1-x-y)\, \Bigg( c_V\, (m_{l_2^+}-m_{l_1^-}) 
\,\eta_i^- + c_A\, (m_{l_2^+}+m_{l_1^-})\, \eta_i^+ \Bigg) \,(\frac{1}
{L^{ver}_{h^0}} - \frac{1}{L^{ver}_{A^0}}) 
\nonumber \\ &+&
(c_V\, \eta_i^A+c_A\,\eta_i^V) 
\Bigg(-2+(q^2\,x\,y-m_{l_1^-}\,m_{l_2^+}\, (-1+x+y)^2) 
(\frac{1}{L^{ver}_{h^0}}+\frac{1}{L^{ver}_{A^0}})-
ln\,\frac{L^{ver}_{h^0}}{\mu^2}\,\frac{L^{ver}_{A^0}}{\mu^2}
\Bigg)
\nonumber \\ &-&
(m_{l_2^+}-m_{l_1^-})\, (1-x-y)\,
\Bigg( \frac{\eta_i^V\,(x\,m_{l_1^-} -y\,m_{l_2^+})+m_i\,\eta_i^+}
{2\,L^{ver}_{A^0\,h^0}}+ \frac{\eta_i^V\,(x\,m_{l_1^-} -y\,m_{l_2^+})-m_i\,
\eta_i^+}{2\,L^{ver}_{h^0\,A^0}}
\Bigg)\nonumber \\
&-& \frac{1}{2} \eta_i^V\, ln\,\frac{L^{ver}_{A^0\,h^0}}{\mu^2}\,
\frac{L^{ver}_{h^0\,A^0}}{\mu^2} 
\Bigg \} \nonumber \,,\\
f_M&=&-\frac{g\, m_W}{64\,\pi^2\,cos\,\theta_W} 
\int_0^1\,dx\, \int_0^{1-x} \, dy \, 
\Bigg \{
\Bigg( (1-x-y)\,(c_V\, \eta_i^V+c_A\,\eta_i^A)\, 
(x\,m_{l_1^-} +y\,m_{l_2^+})
\nonumber \\ &+& \, m_i\,(c_A\, (x-y)\,\eta_i^-+c_V\,\eta_i^+\,(x+y))\Bigg )
\,\frac{1}{L^{ver}_{h^0}}
\nonumber \\ &+& 
\Bigg( (1-x-y)\,
(c_V\, \eta_i^V+c_A\,\eta_i^A)\, (x\,m_{l_1^-} +y\,m_{l_2^+})  
-m_i\,(c_A\, (x-y)\,\eta_i^-+c_V\,\eta_i^+\,(x+y))\Bigg )
\,\frac{1}{L^{ver}_{A^0}}
\nonumber \\ &-& 
(1-x-y) \Bigg (\frac{\eta_i^A\,(x\,m_{l_1^-} +y\,m_{l_2^+})}{2}\, \Big ( 
\frac{1}{L^{ver}_{A^0\,h^0}}+\frac{1}{L^{ver}_{h^0\,A^0}} \Big )
+\frac{m_i\,\eta_i^-} {2} \, \Big ( \frac{1}{L^{ver}_{h^0\,A^0}}-
\frac{1}{L^{ver}_{A^0\,h^0}} \Big ) \Bigg ) \Bigg \} \,,\nonumber \\
f_E&=&-\frac{g\, m_W}{64\,\pi^2\, cos\,\theta_W} 
\int_0^1\,dx\, \int_0^{1-x} \, dy \, 
\Bigg \{
\Bigg( (1-x-y)\,\Big ( -(c_V\, \eta_i^A+c_A\,\eta_i^V)\, (x\,m_{l_1^-} -y\,
m_{l_2^+}) \Big) \nonumber \\ &-&
m_i\, (c_A\, (x-y)\,\eta_i^++c_V\,\eta_i^-\,(x+y))\Bigg )\, 
\frac{1}{L^{ver}_{h^0}}
\nonumber \\ &+&
\Bigg ( (1-x-y)\,\Big ( -(c_V\, \eta_i^A+c_A\,\eta_i^V)\, (x\,m_{l_1^-} -
y\, m_{l_2^+}) \Big ) + m_i\,(c_A\, (x-y)\,\eta_i^++c_V\,\eta_i^-\,(x+y))
\Bigg ) \,\frac{1}{L^{ver}_{A^0}}  
\nonumber \\&+& 
(1-x-y)\, \Bigg ( 
\frac{\eta_i^V}{2}\,(m_{l_1^-}\,x-m_{l_2^+}\, y)\, \, \Big ( 
\frac{1}{L^{ver}_{A^0\,h^0}}+\frac{1}{L^{ver}_{h^0\,A^0}} \Big ) 
+\frac{m_i\,\eta_i^+}{2}\, \Big ( 
\frac{1}{L^{ver}_{A^0\,h^0}}-\frac{1}{L^{ver}_{h^0\,A^0}} \Big )
\Bigg ) 
\Bigg \}\, ,
\label{fVAME}  
\end{eqnarray}
where 
\begin{eqnarray}
L^{self}_{1,\,h^0}&=&m_{h^0}^2\,(1-x)+(m_i^2-m^2_{l_1^-}\,(1-x))\,x
\nonumber \, , \\
L^{self}_{1,\,A^0}&=&L^{self}_{1,\,h^0}(m_{h^0}\rightarrow m_{A^0})
\nonumber \, , \\
L^{self}_{2,\,h^0}&=&L^{self}_{1,\,h^0}(m_{l_1^-}\rightarrow m_{l_2^+})
\nonumber \, , \\
L^{self}_{2,\,A^0}&=&L^{self}_{1,\,A^0}(m_{l_1^-}\rightarrow m_{l_2^+})
\nonumber \, , \\
L^{ver}_{h^0}&=&m_{h^0}^2\,(1-x-y)+m_i^2\,(x+y)-q^2\,x\,y
\nonumber \, , \\
L^{ver}_{h^0\,A^0}&=&m_{A^0}^2\,x+m_i^2\,(1-x-y)+(m_{h^0}^2-q^2\,x)\,y
\nonumber \, , \\
L^{ver}_{A^0}&=&L^{ver}_{h^0}(m_{h^0}\rightarrow m_{A^0})
\nonumber \, , \\
L^{ver}_{A^0\,h^0}&=&L^{ver}_{h^0\,A^0}(m_{h^0}\rightarrow m_{A^0}) \, ,
\label{Lh0A0}
\end{eqnarray}
and 
\begin{eqnarray}
\eta_i^V&=&\xi^{D}_{N,l_1i}\xi^{D\,*}_{N,il_2}+
\xi^{D\,*}_{N,il_1} \xi^{D}_{N,l_2 i} \nonumber \, , \\
\eta_i^A&=&\xi^{D}_{N,l_1i}\xi^{D\,*}_{N,il_2}-
\xi^{D\,*}_{N,il_1} \xi^{D}_{N,l_2 i} \nonumber \, , \\
\eta_i^+&=&\xi^{D\,*}_{N,il_1}\xi^{D\,*}_{N,il_2}+
\xi^{D}_{N,l_1i} \xi^{D}_{N,l_2 i} \nonumber \, , \\
\eta_i^-&=&\xi^{D\,*}_{N,il_1}\xi^{D\,*}_{N,il_2}-
\xi^{D}_{N,l_1i} \xi^{D}_{N,l_2 i}\, . 
\label{etaVA}
\end{eqnarray}
The parameters $c_V$ and $c_A$ are $c_A=-\frac{1}{4}$ and 
$c_V=\frac{1}{4}-sin^2\,\theta_W$. In eq. (\ref{etaVA}) the flavor changing
couplings $\bar{\xi}^{D}_{N, l_ji}$ represent the effective interaction 
between the internal lepton $i$, ($i=e,\mu,\tau$) and outgoing (incoming) 
$j=1\,(j=2)$ one. Here we take  $\bar{\xi}^{D}_{N, l_ji}$ complex in 
general and use the parametrization 
\begin{eqnarray}
\xi^{D}_{N,i l_j}=|\xi^{D}_{N,i l_j}|\, e^{i \theta_{ij}}
\,\, , 
\label{xi}
\end{eqnarray}
where $i,l_j$ denote the lepton flavors and $\theta_{ij}$ are CP violating 
parameters which are the sources of the lepton EDM. 

Now, using the couplings $f_V$, $f_A$, $f_M$ and $f_E$ 
the BR for $Z\rightarrow l_1^-\, l_2^+$ can be written as
\begin{eqnarray}
BR (Z\rightarrow l_1^-\,l_2^+)=\frac{1}{48\,\pi}\,
\frac{m_Z}{\Gamma_Z}\, \{|f_V|^2+|f_A|^2+\frac{1}{2\,cos^2\,\theta_W} 
(|f_M|^2+|f_E|^2) \}
\label{BR1}
\end{eqnarray}
where $\alpha_W=\frac{g^2}{4\,\pi}$ and $\Gamma_Z$ is the total decay width
of Z boson. Note that, in general, the production of sum of charged states 
is considered with the corresponding BR
\begin{eqnarray}
BR (Z\rightarrow l_1^{\pm}\,l_2^{\pm})=
\frac{\Gamma(Z\rightarrow (\bar{l}_1\,l_2+\bar{l}_2\,l_1)}{\Gamma_Z} \, ,
\label{BR2}
\end{eqnarray}
and in our numerical analysis we use this branching ratio. 
\section{Discussion}
In the model III, there are number of free parameters such as the masses of 
charged and neutral Higgs bosons, complex Yukawa couplings. The Yukawa 
couplings in the lepton sector are $\bar{\xi}^D_{N,ij}, i,j=e, \mu, \tau$ 
and they should be restricted by present and forthcoming experiments. The first assumption 
is that $\bar{\xi}^{D}_{N,ij},\, i,j=e,\mu $, are small compared to 
$\bar{\xi}^{D}_{N,\tau\, i}\, i=e,\mu,\tau$ since the strength of these
couplings are related with the masses of leptons denoted by the indices of 
them, similar to the Cheng-Sher scenario \cite{Sher}. Second we assume that 
$\bar{\xi}^{D}_{N,ij}$ is symmetric with respect to the indices $i$ and $j$. 

In our work we study on the decays $Z\rightarrow e^{\pm} \mu^{\pm}$, 
$Z\rightarrow e^{\pm} \tau^{\pm}$ and  $Z\rightarrow \mu^{\pm} \tau^{\pm}$. 
In the case of $Z\rightarrow e^{\pm} \mu^{\pm}$ decay we need the couplings 
$\bar{\xi}^D_{N,\mu i}$ and $\bar{\xi}^D_{N,e i}$ with $i=e,\mu,\tau$. 
Here we use the first asssumption and neglect the contributions of 
$\bar{\xi}^D_{N,i j}$ where $i,j=\mu,e$, by taking only the internal $\tau$ 
lepton into account. Now, we should restrict $\bar{\xi}^D_{N,i \tau}$, 
$i=e,\mu$. For $\bar{\xi}^D_{N,\mu \tau}$ we use the constraint coming from 
the experimental limits of $\mu$ lepton EDM \cite{Abdullah},
\begin{eqnarray}
0.3\times 10^{-19}\, e-cm < d_{\mu} < 7.1\times 10^{-19}\, e-cm 
\label{muedmex}
\end{eqnarray}
(see \cite{ErLFV} for details).  

For the restriction of the coupling $\bar{\xi}^D_{N,\mu \tau}$, the
deviation of the anomalous magnetic moment (AMM) of muon over its SM 
prediction \cite{Czarnecki} due to the the recent experimental result 
of muon AMM by g-2 Collaboration \cite{Brown}, can also be used. However, 
AMM of muon is posssible in the model III even for vanishing complex Yukawa 
couplings. In the LFV $Z\rightarrow l_1 l_2$ decay, the part which depends 
on the couplings $\eta_i^A$ and $\eta_i^-$ (see eq. (\ref{etaVA}) ) is 
non-vanishing when the complex Yukawa couplings are permitted in the model. 
Here, we choose EDM of muon since the restriction is sensitive to complex 
phases, existing also in the $Z\rightarrow l_1 l_2$ decay. The coupling 
$\bar{\xi}^D_{N,e \tau}$ is restricted using the experimental upper limit 
of the $BR$ of the process $\mu\rightarrow e\gamma$ and the above constraint 
for $\bar{\xi}^D_{N,\mu \tau}$, since $\mu\rightarrow e\gamma$ decay can be 
used to fix the Yukawa combination $\bar{\xi}^{D}_{N,\mu\tau}\,
\bar{\xi}^{D}_{N,e\tau}$ (see \cite{ErLFV}). This ensures us to determine 
the upper and lower limits of the coupling $\bar{\xi}^D_{N,e \tau}$. The 
maximum value of the BR($Z\rightarrow \mu \,e$) is calculated by taking 
the combination $\bar{\xi}_{N,\mu\tau}^D \bar{\xi}_{N,e\tau}^D$, which 
respects the upper bound of $\mu\rightarrow e\gamma$ decay. For the minimum 
value of BR($Z\rightarrow \mu\, e$), we use the combination 
$\bar{\xi}_{N,\mu\tau}^D \bar{\xi}_{N,e\tau}^D$ if each coupling is at its 
minimum value. In fact, this minimum value is artificial. Note that, 
$\bar{\xi}^D_{N,e \tau}$ can be restricted and the minimum value of 
BR($Z\rightarrow \mu e$) can be obtained by using the experimental 
result of the EDM of electron \cite{Commins}. However, we expect that the 
experimental result of the EDM of electron is not more reliable than the 
one of the process $\mu\rightarrow e\gamma$.

This analysis shows that $|\bar{\xi}^{D}_{N,\mu\tau}|$, 
($|\bar{\xi}^{D}_{N,e\tau}|$) is at the order
of the magnitude of $10^3$ ($10^{-4}$) GeV. Note that these couplings are
chosen complex to be able to describe the EDM which is posssible in the
case of CP violating interactions. Throughout our calculations we use the 
input values given in Table (\ref{input}).  
\begin{table}[h]
        \begin{center}
        \begin{tabular}{|l|l|}
        \hline
        \multicolumn{1}{|c|}{Parameter} & 
                \multicolumn{1}{|c|}{Value}     \\
        \hline \hline
        $m_{\mu}$                   & $0.106$ (GeV) \\
        $m_{\tau}$                   & $1.78$ (GeV) \\           
        $m_{W}$             & $80.26$ (GeV) \\
        $m_{Z}$             & $91.19$ (GeV) \\
        $G_F$             & $1.16637 10^{-5} (GeV^{-2})$  \\
        $\Gamma_Z$                  & $2.490\, (GeV)$  \\
        $sin\,\theta_W$               & $\sqrt{0.2325}$ \\
        \hline
        \end{tabular}
        \end{center}
\caption{The values of the input parameters used in the numerical
          calculations.}
\label{input}
\end{table}

Fig. \ref{BrZemuMaxsintet} (\ref{BrZemuMinsintet}) represents 
$sin\theta_{\tau e}$ dependence of the maximum (minimum) value of the 
$BR\,(Z\rightarrow \mu^{\pm}\, e^{\pm})$ for $sin\,\theta_{\tau \mu}=0.5$, 
$m_{h^0}=70\, GeV$ and $m_{A^0}=80\, GeV$. Here the maximum and minimum 
values are predicted by taking upper and lower limits of $\mu$ lepton 
EDM into account. The maximum (minimum) value of the BR is $7\times 10^{-11}$ 
($10^{-13}$) for small values of $sin\theta_{\tau e}$ and its sensivity to 
this parameter is weak. $BR$ decreases at the order of the magnitude $15 \% $ 
for $sin\theta_{\tau e} \ge 0.5$ and it becomes more sensitive to 
$sin\theta_{\tau e}$. $sin\theta_{\tau \mu}$ dependence of the maximum 
(minimum) value of the $BR\,(Z\rightarrow \mu^{\pm}\, e^{\pm})$ for 
$sin\theta_{\tau e}=0.5$, $m_{h^0}=70\, GeV$ and $m_{A^0}=80\, GeV$ almost
the same as $sin\theta_{\tau e}$ dependence of the $BR$ under consideration.

In Fig. \ref{BrZemuMinA0} we present $m_{A^0}$ dependence of the 
minimum value of the $BR\,(Z\rightarrow \mu^{\pm}\, e^{\pm})$ for 
$sin\theta_{\tau e}=0.5$, $sin\theta_{\tau \mu}=0.5$ and $m_{h^0}=70\, GeV$. 
The $BR$ is strongly sensitive to $m_{A^0}$ and decreases with increasing
values of $m_{A^0}$. The same dependence appears for the maximum value of
the $BR$.

In the $Z\rightarrow \tau^{\pm} e^{\pm}$ decay, the couplings 
$\bar{\xi}^D_{N,\tau i}$ and $\bar{\xi}^D_{N,e i}$ with $i=e,\mu,\tau$ play 
the main role, in the model III. Again the first assumption permits us 
to neglect the contributions of the couplings $\bar{\xi}^D_{N,i j}$ where 
$i,j=\mu,e$, by taking only the internal $\tau$ lepton into account. For the 
coupling $\bar{\xi}^{D}_{N,e\tau}$ the same restriction is used, however for
$\bar{\xi}^{D}_{N,\tau\tau}$ we do not use any constraint. 
 
Fig. \ref{BrZetauMaxsintet} (\ref{BrZetauMinsintet}) shows 
$sin\theta_{\tau e}$ and $sin\theta_{\tau \mu}$ dependence of the maximum 
(minimum) value of the $BR\,(Z\rightarrow \tau^{\pm}\, e^{\pm})$ for 
$sin\theta_{\tau \mu}=0.5$ and $sin\theta_{\tau e}=0.5$ respectively. 
Here the coupling $\bar{\xi}^{D}_{N,\tau\tau}$ is taken as 
$\bar{\xi}^{D}_{N,\tau\tau}=10^3\, GeV$. The maximum (minimum) value of the 
$BR$ for this process is at the order of the magnitude of $10^{-11}$ 
$(10^{-12})$. $BR$ increases with increasing values $sin\theta_{\tau \mu}$,
however this behaviour appears in reverse for $sin\theta_{\tau e}$ 
dependence. The sensitivity of $BR\,(Z\rightarrow \tau^{\pm}\, e^{\pm})$ to 
both CP violating parameters, $sin\theta_{\tau e}$ and $sin\theta_{\tau
\mu}$, is strong. 

We present the coupling $\bar{\xi}^{D}_{N,\tau\tau}$ dependence of 
$BR\,(Z\rightarrow \tau^{\pm}\, e^{\pm})$ in Figs. \ref{BrZetauMaxksiDtautau} 
and \ref{BrZetauMinksiDtautau}. These figures shows that the $BR$ enormously 
increases with increasing values of the coupling
$\bar{\xi}^{D}_{N,\tau\tau}$. Note that we take the coupling
$\bar{\xi}^{D}_{N,\tau\tau}$ real in our calculations.

Finally, we study the $Z\rightarrow \mu^{\pm} \tau^{\pm}$ decay by taking 
$\tau$ lepton as an internal one similar to previous analysis. Here the 
experimental upper limit for BR($Z\rightarrow \mu^{\pm} \tau^{\pm}$) can 
be reached for the small values of $\bar{\xi}^{D}_{N,\tau\tau}$. Further, 
the theoretical calculations are consistent with this upper limit for 
the case where the mass differences of neutral Higgs bosons $h^0$ and $A^0$ 
are large, even for large values of the coupling $\bar{\xi}^{D}_{N,\tau
\tau}$. 

As a summary, we study the $BR$'s of the decays 
$Z\rightarrow e^{\pm} \mu^{\pm}$, $Z\rightarrow e^{\pm} \tau^{\pm}$ and  
$Z\rightarrow \mu^{\pm} \tau^{\pm}$ and observe that it is
possible to reach the present experimental upper limits in the model III, 
playing with the model parameters in the restriction region. This result is
important since the theoretical work in the SM shows that the branching
rates are less than $10^{-54}$ and compared to this number large rates are 
expected with massive and mixing neutrinos. In our analysis, we predict 
that the $BR$ for the $Z\rightarrow e^{\pm} \mu^{\pm}$ decay can reach to 
the values at the order of the magnitude $10^{-11}$. $BR$ for the process 
$Z\rightarrow e^{\pm} \tau^{\pm}$ depends strongly on the Yukawa coupling 
$\bar{\xi}^{D}_{N,\tau\tau}$ and for its large values, $10^3-10^4\, GeV$, 
it can be in the range $10^{-10}-10^{-9}$. The process 
$Z\rightarrow \mu^{\pm} \tau^{\pm}$ can have larger $BR$ compared to the
previous ones, since the Yukawa couplings entering in the expressions 
are $\bar{\xi}^{D}_{N,\mu\tau}$ and $\bar{\xi}^{D}_{N,\tau\tau}$. 
Furthermore, $BR$'s of the processes under consideration are sensitive to
the CP-violating parameters since the source of the parts which depend on 
couplings $\eta_i^A$ and $\eta_i^-$ are the non-vanishing complex Yukawa 
couplings, in the model III. 

In future, with the reliable experimental result of upper limits of the 
$BR$'s of above processes it would be possible to test models beyond the 
SM and free parameters of these models
\section{Acknowledgement} The work of E.O.I was supported by Turkish Academy
of Sciences (T\"UBA/GEBIP).

\newpage
\begin{figure}[htb]
\vskip 0.0truein
\centering
\epsfxsize=6.0in
\leavevmode\epsffile{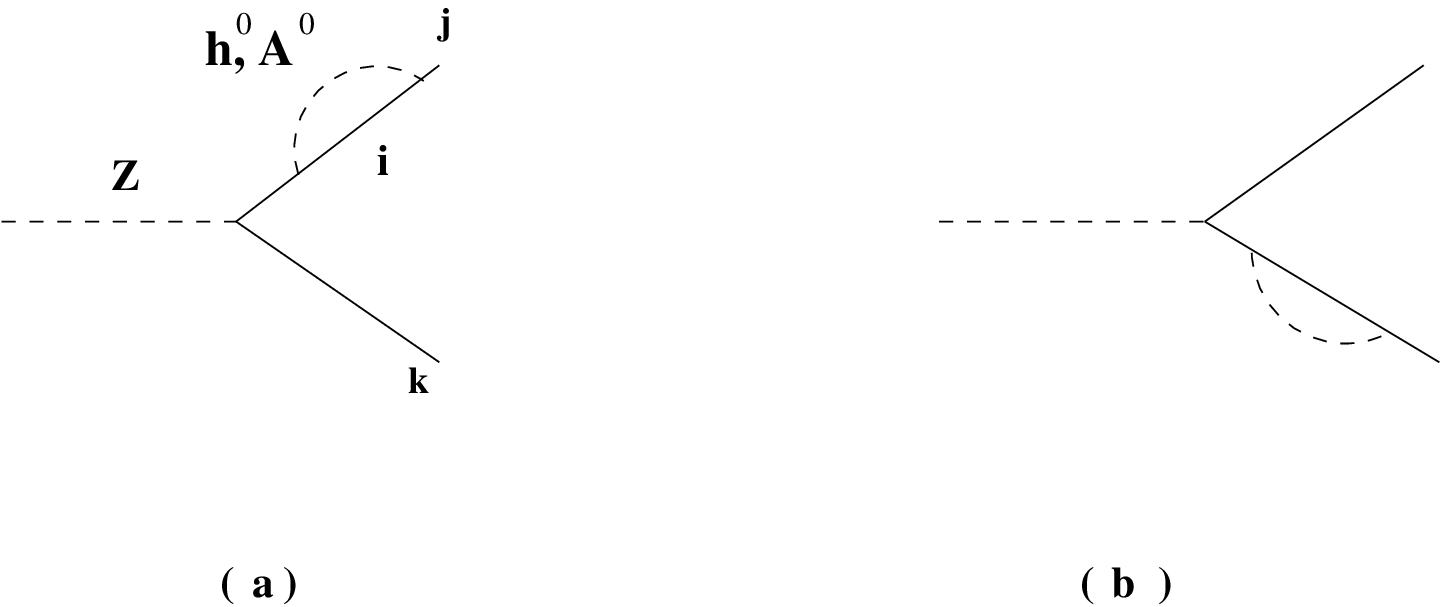}
\vskip 0.5truein
\end{figure}
\begin{figure}[htb]
\vskip 0.0truein
\centering
\epsfxsize=6.0in
\leavevmode\epsffile{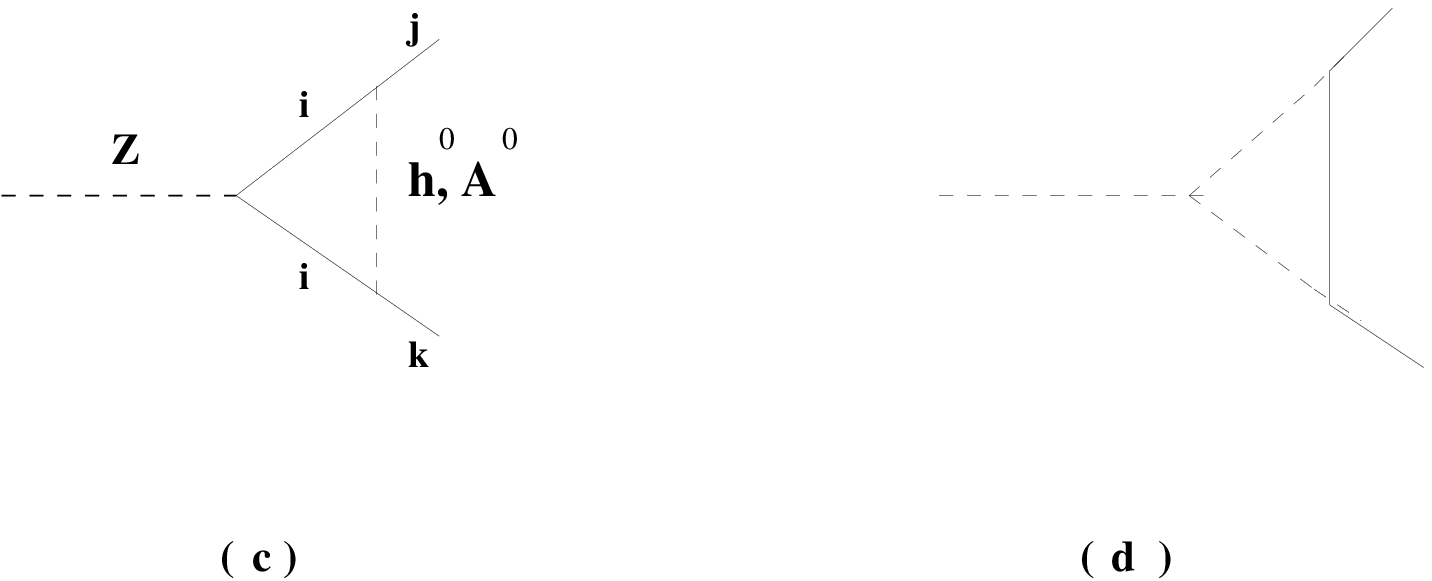}
\vskip 0.5truein
\caption[]{One loop diagrams contribute to $Z\rightarrow k^+\,j^-$ decay 
due to the neutral Higgs bosons $h_0$ and $A_0$ in the 2HDM. $i$ represents
the internal, $j$ ($k$) outgoing (incoming) lepton, dashed lines the vector 
field Z, $h_0$ and $A_0$ fields.}
\label{fig1ver}
\end{figure}
\newpage
\begin{figure}[htb]
\vskip -3.0truein
\centering
\epsfxsize=6.8in
\leavevmode\epsffile{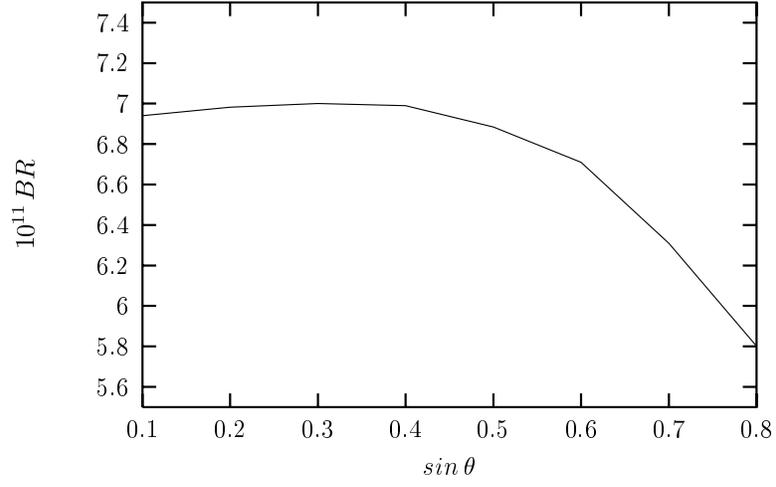}
\vskip -3.0truein
\caption[]{The maximum value of $BR\,(Z\rightarrow \mu^{\pm}\, e^{\pm})$ as 
a function of $sin\theta_{\tau e}$ for $sin\theta_{\tau \mu}=0.5$, 
$m_{h^0}=70\, GeV$ and $m_{A^0}=80\, GeV$. } 
\label{BrZemuMaxsintet}
\end{figure}
\begin{figure}[htb]
\vskip -3.0truein
\centering
\epsfxsize=6.8in
\leavevmode\epsffile{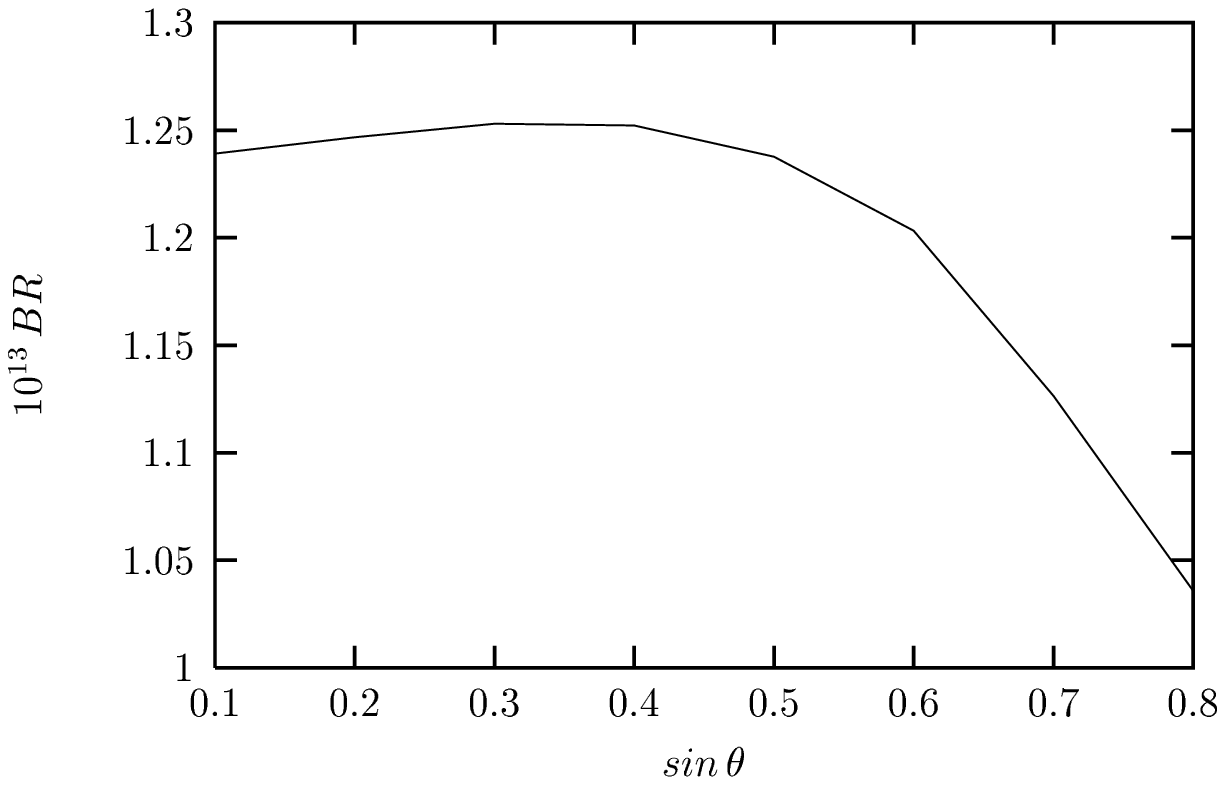}
\vskip -3.0truein
\caption[]{The same as Fig. \ref{BrZemuMaxsintet} but for 
the minimum value of $BR\,(Z\rightarrow \mu^{\pm}\, e^{\pm})$ .}
\label{BrZemuMinsintet}
\end{figure}
\begin{figure}[htb]
\vskip -3.0truein
\centering
\epsfxsize=6.8in
\leavevmode\epsffile{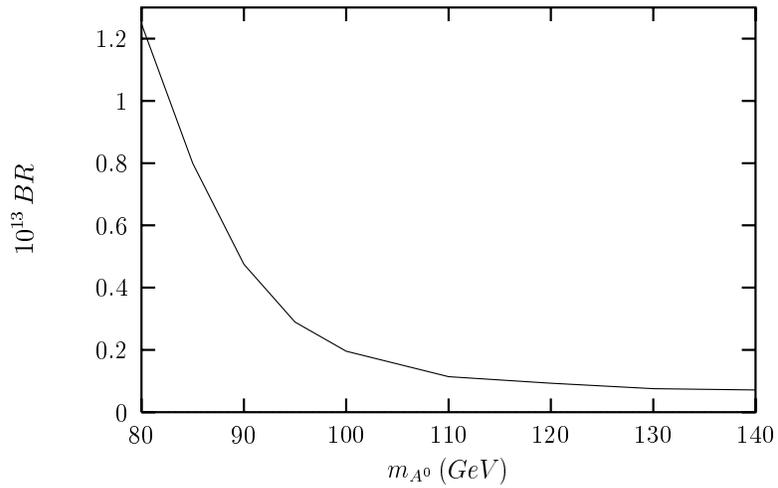}
\vskip -3.0truein
\caption[]{The minimum value of $BR\,(Z\rightarrow \mu^{\pm}\, e^{\pm})$ 
as a function of $m_{A^0}$ for $sin\theta_{\tau \mu}=0.5$, $sin\theta_
{\tau e}=0.5$ and $m_{h^0}=70\, GeV$. }
\label{BrZemuMinA0}
\end{figure}
\begin{figure}[htb]
\vskip -3.0truein
\centering
\epsfxsize=6.8in
\leavevmode\epsffile{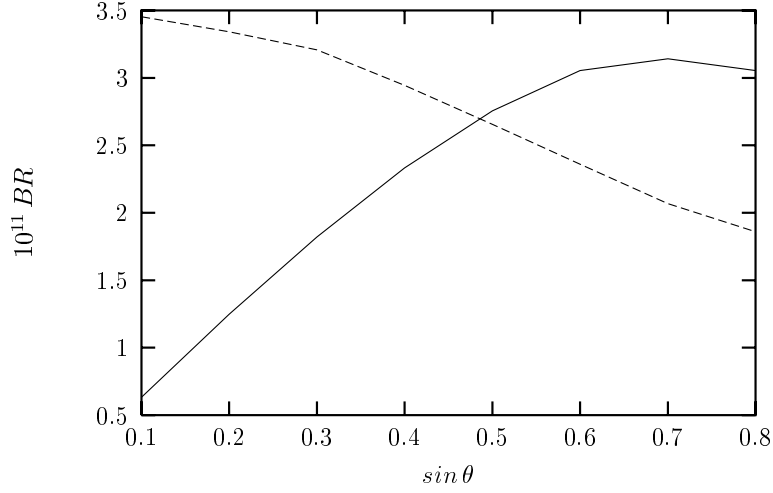}
\vskip -3.0truein
\caption[]{The maximum value of $BR\,(Z\rightarrow \tau^{\pm}\, e^{\pm})$ 
as a function of $sin\theta$ for $\bar{\xi}^{D}_{N,\tau\tau}=10^3\, GeV$, 
$m_{h^0}=70\, GeV$ and $m_{A^0}=80\, GeV$. Here solid line represents the 
dependence with respect to $sin\theta_{\tau \mu}$ for $sin\theta_{\tau e}
=0.5$ and dashed line to $sin\theta_{\tau e}$ for 
$sin\theta_{\tau \mu}=0.5$.} 
\label{BrZetauMaxsintet}
\end{figure}
\begin{figure}[htb]
\vskip -3.0truein
\centering
\epsfxsize=6.8in
\leavevmode\epsffile{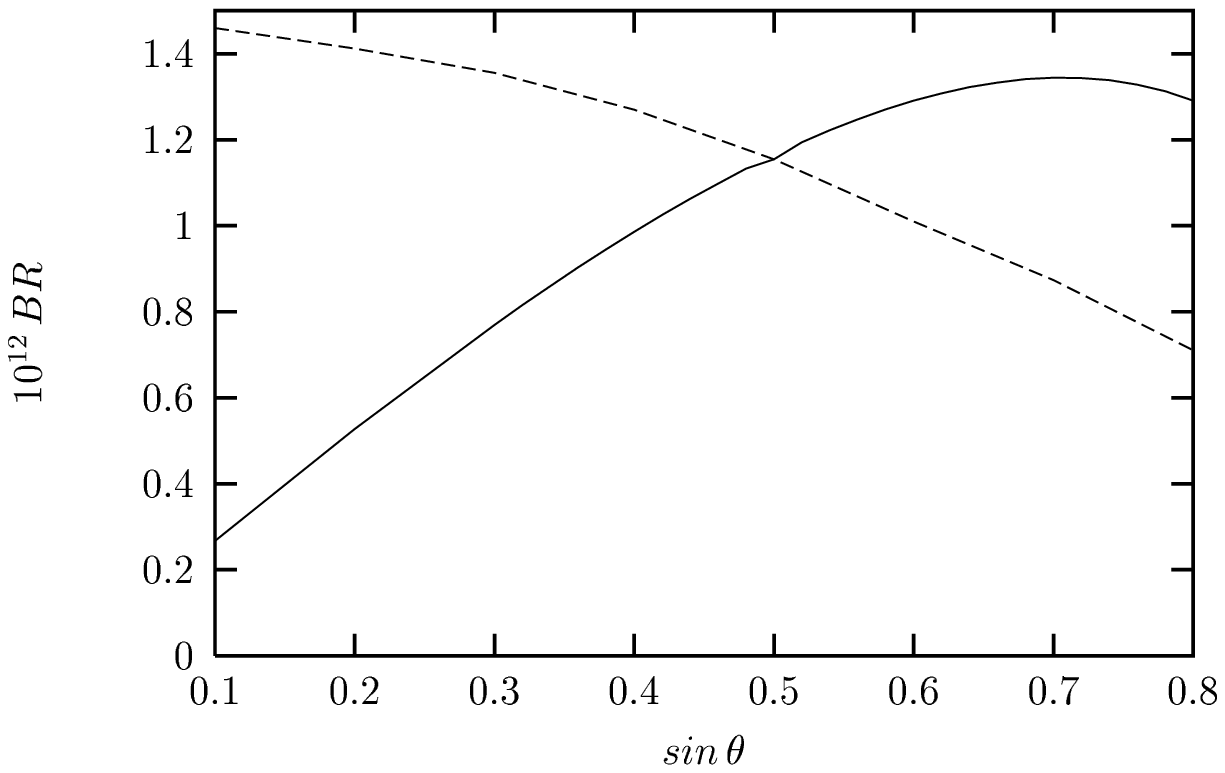}
\vskip -3.0truein
\caption[]{The same as Fig. \ref{BrZetauMaxsintet} but for 
the minimum value of $BR\,(Z\rightarrow \tau^{\pm}\, e^{\pm})$. }
\label{BrZetauMinsintet}
\end{figure}
\begin{figure}[htb]
\vskip -3.0truein
\centering
\epsfxsize=6.8in
\leavevmode\epsffile{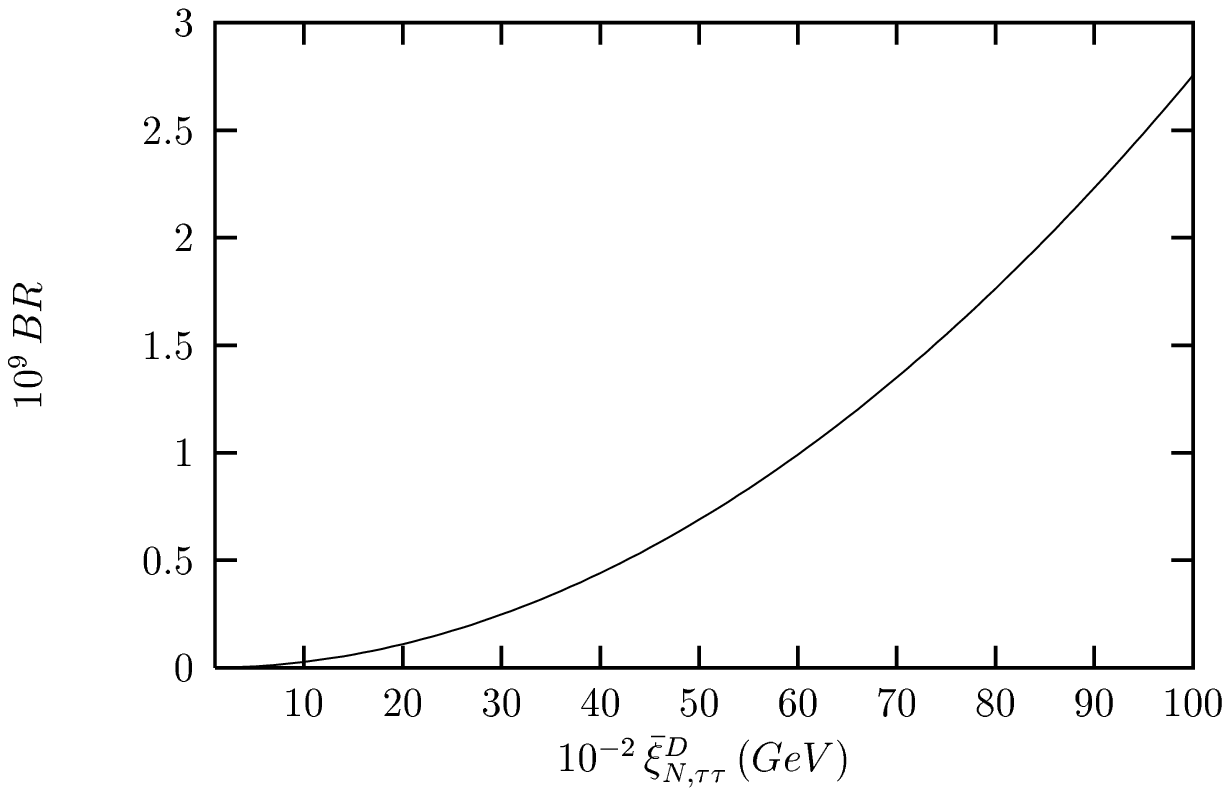}
\vskip -3.0truein
\caption[]{The maximum value of $BR\,(Z\rightarrow \tau^{\pm}\, e^{\pm})$ 
as a function of $\bar{\xi}^D_{N,\tau\tau}$ for $sin\theta_{\tau \mu}=0.5$, 
$sin\theta_{\tau e}=0.5$, $m_{h^0}=70\, GeV$ and $m_{A^0}=80\, GeV$. } 
\label{BrZetauMaxksiDtautau}
\end{figure}
\begin{figure}[htb]
\vskip -3.0truein
\centering
\epsfxsize=6.8in
\leavevmode\epsffile{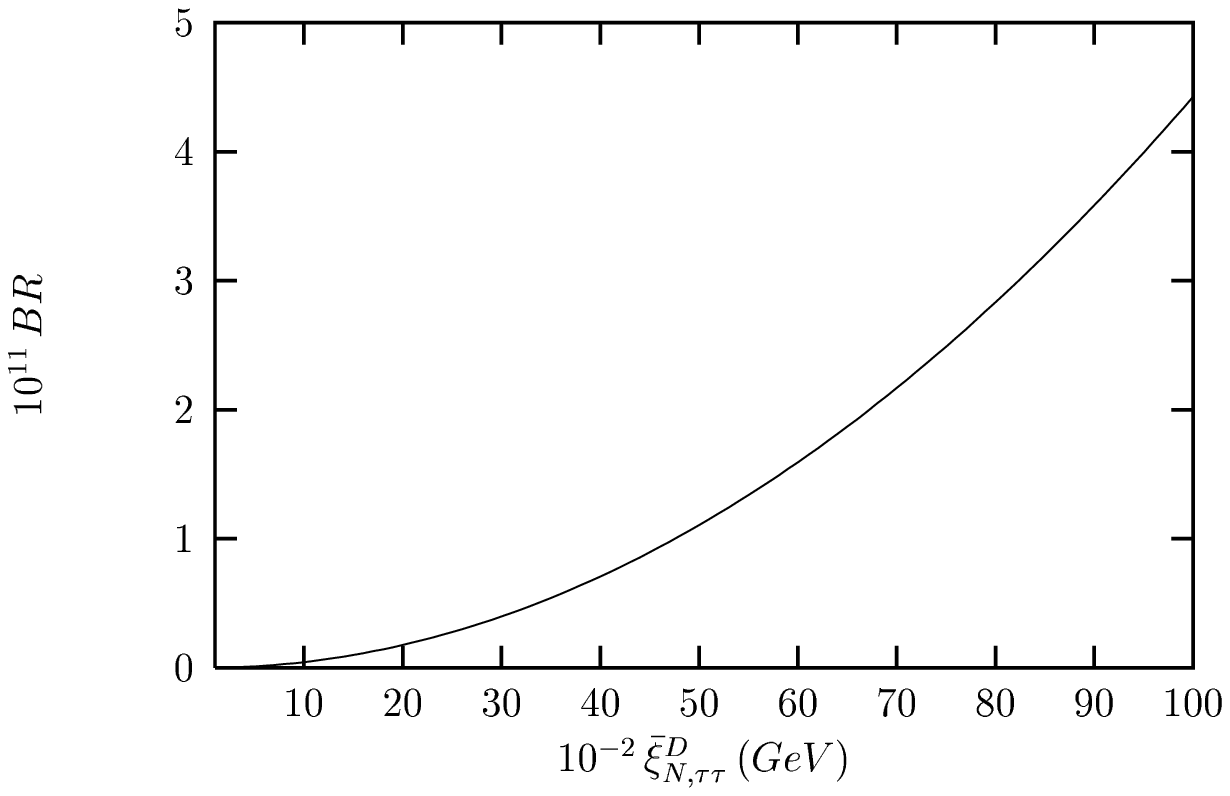}
\vskip -3.0truein
\caption[]{The same as Fig. \ref{BrZetauMaxksiDtautau} but for 
the minimum value of $BR\,(Z\rightarrow \tau^{\pm}\, e^{\pm})$. }
\label{BrZetauMinksiDtautau}
\end{figure}
\end{document}